\documentclass[conference]{IEEEtran}
\IEEEoverridecommandlockouts
\usepackage{cite}
\usepackage{amsmath,amssymb,amsfonts}
\usepackage{algorithmic}
\usepackage{graphicx}
\usepackage{textcomp}
\usepackage{xcolor}
\usepackage{subfig}
\usepackage{epstopdf}
\def\BibTeX{{\rm B\kern-.05em{\sc i\kern-.025em b}\kern-.08em
    T\kern-.1667em\lower.7ex\hbox{E}\kern-.125emX}}
\begin{document}

\title{GAN-based Hyperspectral Anomaly Detection \\
\thanks{This work is supported by Scientific and Technological Research Council of Turkey (TUBITAK) under Project No. 118E295. The work of S. Arisoy was supported by Council of Higher Education (YOK) of Turkey through the YUDAB scholarship program.}
}

\author{\IEEEauthorblockN{Sertac Arisoy}
\IEEEauthorblockA{\textit{Electronics Engineering Dept.} \\
\textit{Gebze Technical University}\\
Kocaeli, Turkey \\
sarisoy@gtu.edu.tr}
\and
\IEEEauthorblockN{Nasser M. Nasrabadi}
\IEEEauthorblockA{\textit{Lane Dept. of Computer Sci. and Electrical Engin.} \\
\textit{West Virginia University}\\
Morgantown, WV, USA\\
nasser.nasrabadi@mail.wvu.edu}
\and
\IEEEauthorblockN{Koray Kayabol}
\IEEEauthorblockA{\textit{Electronics Engineering Dept.} \\
\textit{Gebze Technical University}\\
Kocaeli, Turkey \\
koray.kayabol@gtu.edu.tr}
}

\maketitle

\begin{abstract}
In this paper, we propose a generative adversarial network (GAN)-based hyperspectral anomaly detection algorithm. In the proposed algorithm, we train a GAN model to generate a synthetic background image which is close to the original background image as much as possible. By subtracting the synthetic image from the original one, we are able to remove the background from the hyperspectral image. Anomaly detection is performed by applying Reed-Xiaoli (RX) anomaly detector (AD) on the spectral difference image. In the experimental part, we compare our proposed method with the classical RX, Weighted-RX (WRX) and support vector data description (SVDD)-based anomaly detectors and deep autoencoder anomaly detection (DAEAD) method on synthetic and real hyperspectral images. The detection results show that our proposed algorithm outperforms the other methods in the benchmark. 
\end{abstract}


\section{Introduction}
Hyperspectral sensors collect rich information from the scene by hundreds of spectral bands. The rich spectral information makes the classification of image scene or the detection of target or anomaly pixels be possible  \cite{matteoli2010tutorial}. Anomaly detection in hyperspectral image is used to detect the rare pixels which have different spectral signatures compared to the background pixels. The Reed-Xiaoli (RX) \cite{RX90} anomaly detector (AD) is the fundamental benchmarking algorithm in hyperspectral anomaly detection. The RX-AD characterizes the background as a multivariate Gaussian distribution and assigns anomaly scores according to the Mahalanobis distance. Although the RX-AD is simple and more practical for real applications, it has some limitations. Different versions of the RX algorithm have been proposed such as the Regularized-RX \cite{nasrabadi2008regularization}, Kernel-RX \cite{KwonNasrabadi05}, Weighted-RX (WRX) \cite{guo2014weighted}. In the regularized-RX-AD \cite{nasrabadi2008regularization}, the covariance matrix is regularized with the identity matrix to overcome the singularity problem caused by using a small sample size in the estimation. In the Kernel-RX-AD \cite{KwonNasrabadi05}, the pixels are projected onto a high dimensional feature space by a nonlinear function and anomaly detection is performed in this space. In the WRX-AD \cite{guo2014weighted}, the impact of the anomaly pixels during the background parameter estimation is reduced by assigning some weights to the pixels. There are also deterministic detectors in the literature such as the support vector data description (SVDD) where the background model is created with an enclosed hypersphere and anomaly score is assigned to the test pixel according to the distance from the hypersphere center.

Recently, deep learning methods have been adopted for hyperspectral anomaly detection. In \cite{bati2015hyperspectral}, an auto-encoder model is proposed for hyperspectral image reconstruction and the reconstruction error values are used for anomaly assignment. In \cite{taghipour2019unsupervised}, similarly, a deep autoencoder anomaly detection (DAEAD) method is proposed and the anomaly score is calculated from the reconstruction error. In \cite{ma2018unsupervised}, a deep belief network (DBN) is used for the feature representation and input reconstruction. At the test phase, the representative features are generated from DBN and the anomaly impact is reduced with adaptive weights obtained from the reconstruction error. Anomaly score is calculated by a linear combination of the representative feature similarities between the test pixel and the neighbors within the local area according to adaptive weights. In \cite{li2017transferred}, the spectral similarity measurement between two vectors is learned by a convolutional neural network (CNN) using a reference labeled image. The anomaly score is calculated by averaging the similarity scores between the test pixel and the neighbors within a local area. In \cite{lei2019spectral}, two separate detection maps are found in the spectral and spatial domains. In the spectral domain, a DBN is used for dimensionality reduction and the anomaly score is calculated with the RX-AD. In the spatial domain, a morphological attribute filter (MAF) is used to detect small objects in the image. The final detection map is obtained with the combination of these two maps. The generative adversarial networks (GAN) also has been utilized for hyperspectral anomaly detection. In \cite{eide2018applying}, Wasserstein's GAN model is proposed for hyperspectral anomaly detection. The generator- and discriminator-based detectors are used for the detection. In \cite{xie2019spectral}, the background is suppressed with an adversarial autoencoder (AEE) model and an MAF. The representative features are learned by AEE and combined for the initial map. An MAF is applied to the map and the filtered map is used for background suppression. The anomaly score is calculated with the Mahalanobis distance. 

\begin{figure*}[h]
	\centering{
		\includegraphics[width=17cm]{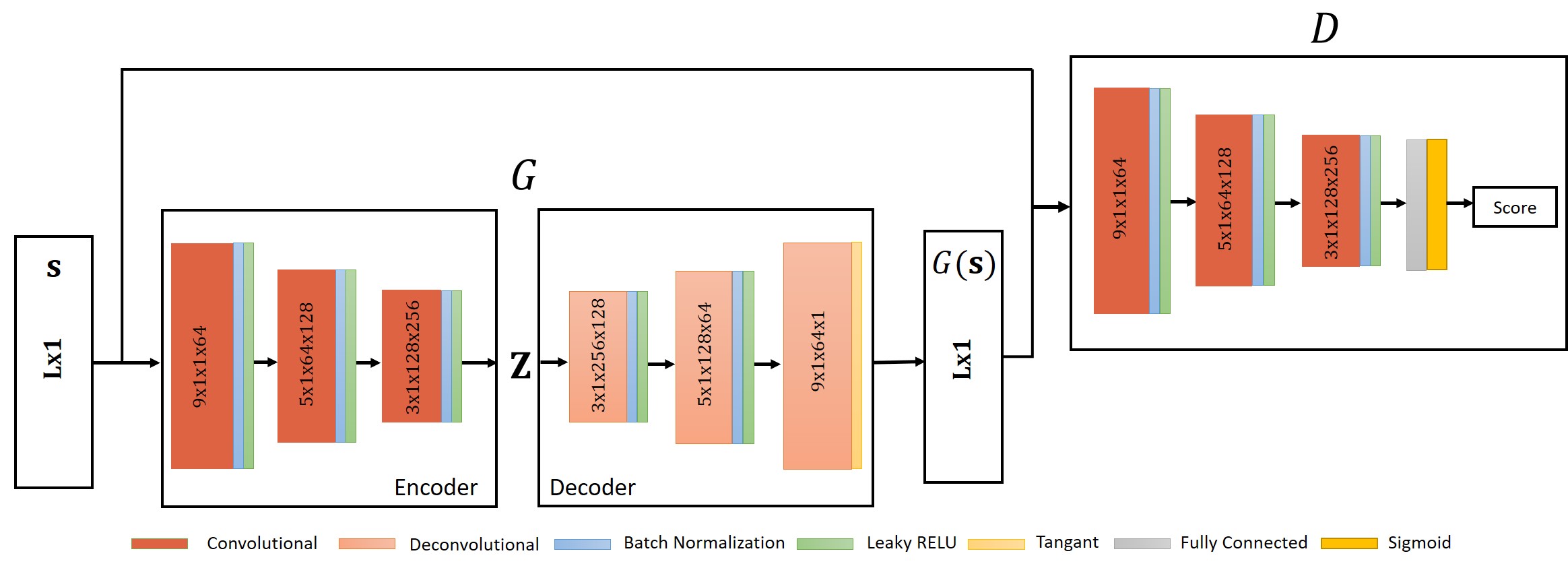}		
	}%
	\caption{Proposed conditional GAN model. Layer parameters indicate: kernel width, kernel height, input channel number, and output channel number}
	\label{ganmodel}
\end{figure*}

Different from those papers, in this paper, we propose a conditional GAN model to generate a synthetic hyperspectral image. The GAN model learns the input data statistics with a cross-entropy function and generates fake data by using the generator. We use the synthetic hyperspectral image to suppress the background parts of the image. For this purpose, we subtract the synthetic image from the original image and obtain a spectral difference image. We calculate the anomaly scores with the RX-AD in the difference image. Our proposed method is called GAN-RX.

The organization of the paper is as follows. Section \ref{SecProposed} explains the GAN model and the proposed detector. The experimental results are presented in Section \ref{SecExp}. Section \ref{SecConc} summarizes the proposed method and explains the future works.

\section{Proposed Method}
\label{SecProposed}

We propose to use the GAN model shown in Fig. \ref{ganmodel} to reconstruct the spectral vectors. Since the number of background pixels is higher than the number of anomaly pixels, the generator learns the background characteristics better. Consequently, the anomaly pixels are reconstructed with a higher error than the background pixels. We subtract the reconstructed image from the original image to obtain a spectral difference image. After the subtraction of the generated background by GAN, we expect that the majority parts of the background are removed in the spectral difference image. We calculate the anomaly score by applying the RX-AD to the difference image. We first summarize the proposed GAN model and then explain the detector in the following two sub-sections.

\subsection{Reconstruction Step: GAN}

The GAN model is first introduced in \cite{goodfellow2014generative}. The model consists of two sub-networks called generator and discriminator. The generator sub-network learns the input data statistics by following the discriminator output score and generates the fake data to be as similar to the input data $\mathbf{s} \sim f_{\mathbf{S}}(\mathbf{s})$ as possible by projecting the input noise vector $\mathbf{z} \sim f_{\mathbf{Z}}(\mathbf{z})$. The discriminator sub-network learns to distinguish whether the input data is real or fake, and generates a probability score at the output. In \cite{radford2015unsupervised}, a deep convolutional GAN model is proposed such that the sub-networks are composed of convolutional layers. In \cite{akcay2018ganomaly} and \cite{sabokrou2018adversarially}, the GAN model is conditioned on the input data to generate also the input noise vector $\mathbf{z}$ from the data. In this model the input noise vector of classical GAN is called representative features and corresponds to the latent space or middle layer of an encoder-decoder sub-network. In \cite{akcay2018ganomaly}, the proposed GAN model is used for anomaly detection in MNIST \cite{lecun2010mnist} and CIFAR \cite{krizhevsky2009learning} datasets by choosing one of the classes as an anomaly class and assuming the others as a single normal class. In \cite{sabokrou2018adversarially}, the GAN model is proposed for novelty detection as a one-class classifier. The model is tested for outlier images detection in datasets and anomaly detection in videos. In this paper, we adopt this model for hyperspectral anomaly detection. In our proposed model, the generator sub-network composes of encoder and decoder parts as shown in Fig. \ref{ganmodel}. The encoder part includes three convolutional layers and each convolutional layer is followed by batch-normalization and leaky ReLU operators. The decoder part includes three deconvolutinal layers and the first and the second layers are followed by batch-normalization and leaky ReLU operators. The third layer is followed by a hyperbolic tangent operator. The discriminator sub-network is similar to the encoder and it generates a probability score at the output with a logistic sigmoid function. The weights of the generator and the discriminator networks are trained according to the following objective function: 

\begin{equation}
\begin{split}
\mathcal{L}_{GAN}(G,D)=\mathrm{E}_{\mathbf{s}} [\log(D(\mathbf{s}))] + \\
                        \mathrm{E}_{\mathbf{s}} [\log(1-D(G(\mathbf{s})))],
\end{split}
 \label{eq:gan}
\end{equation}
where $\mathbf{s} \in \mathbb{R}^{L}$ is the real spectral vector. We also add a reconstruction loss constraint to the objective function in order to reconstruct the input samples with a low error rate. The reconstruction loss function is given as  $ \mathcal{L}_\mathcal{R} = \mathrm{E}_{\mathbf{s}} \left[ \|  \mathbf{s}-G(\mathbf{s})\|_1 \right]$. The total loss function is defined as follows: 

\begin{equation}
  \mathcal{L}_{total}=\mathcal{L}_{GAN}(G,D)+ \alpha \mathcal{L}_\mathcal{R},
\label{eq:gantotal}
\end{equation}
where the parameter $\alpha$ balances the two terms in (\ref{eq:gantotal}).

\subsection{The Detection Step: GAN-RX}

After the training of sub-networks weights, the generator is able to reconstruct the input sample with a low error. We expect that the spectral vectors of the background pixels are reconstructed with lower error than those of anomaly pixels. We remove the background by subtracting the reconstructed image from the original one.  The $i$th pixel of the spectral difference image is defined by

\begin{equation}
\label{csmp12}
\mathbf{d}_{i} = \mathbf{s}_i-G(\mathbf{s}_i),
\end{equation}
where $i = 1, \dots , N$ is the pixel indices and $G(\mathbf{s}_i)$ is the reconstructed spectral vector. The mean vector $\mathbf{m}$ and the covariance matrix $\mathbf{C}$ are estimated from the difference image by the following equations:

\begin{equation}\label{mean}
\hat{\mathbf{m}} = \sum_{i=1}^{N}\mathbf{d}_i,
\end{equation}

\begin{equation}\label{cov}
\hat{\mathbf{C}} =\sum_{i=1}^{N}(\mathbf{d}_i-\hat{\mathbf{m}} )(\mathbf{d}_i-\hat{\mathbf{m}} )^{T}.
\end{equation}
The anomaly score is calculated for $i$th pixel as follows:

\begin{equation}\label{detection}
\mathrm{GAN-RX}(\mathbf{d}_i)=(\mathbf{d}_i-\hat{\mathbf{m}} )^{T}\hat{\mathbf{C}}^{-1}(\mathbf{d}_i-\hat{\mathbf{m}} ).
\end{equation}

\begin{figure}[t]
	\centering{
		\includegraphics[width=7cm]{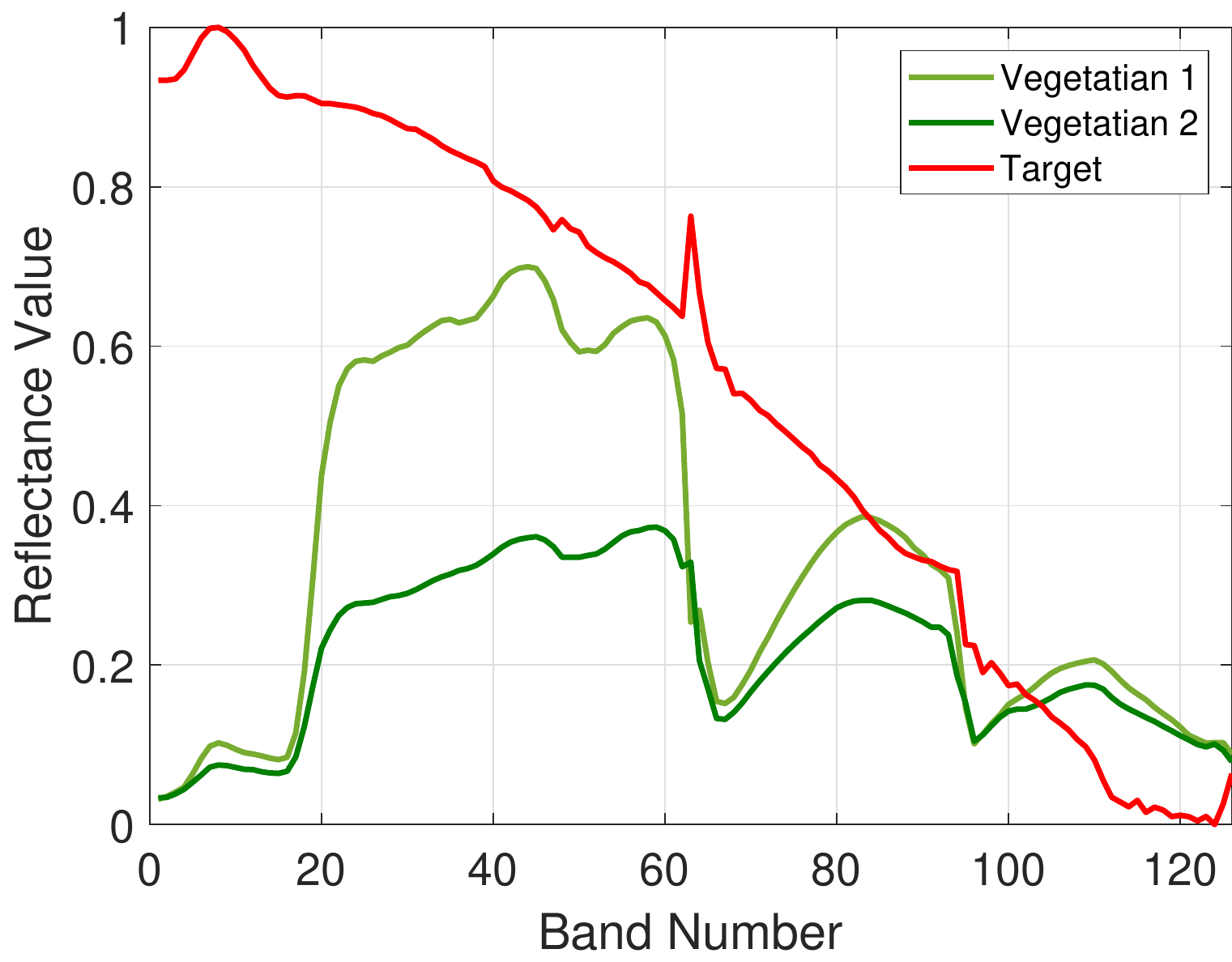}		
	}%
	\caption{Spectral signatures of implant target and vegetation backgrounds.}
	\label{spectrumss}
\end{figure}

\section{Experimentral Results}
\label{SecExp}

In this section, we evaluate our proposed method on synthetic and real hyperspectral images. We use the target implant method \cite{stefanou2009method} for synthetic image. We measure the detection performances with receiver operating characteristics (ROC) curves and the area under the curves (AUC). We compare our proposed method with the classical RX \cite{RX90}, WRX \cite{guo2014weighted} and SVDD-based \cite{banerjee2006support} anomaly detectors and  DAEAD \cite{taghipour2019unsupervised} method.

\begin{table}
		\renewcommand{\arraystretch}{1.4}
		\begin{tabular}{p{2.45cm}p{0.63cm}p{0.63cm}p{0.63cm}p{0.77cm}p{1.15cm}}
			\hline
			\textbf{Images}    & \textbf{RX}  & \textbf{WRX}  &\textbf{SVDD}  &\textbf{DAEAD} &\textbf{GAN-RX} \\
			Cooky City Synthetic & 0.3028 & 0.2944 & 0.9461 & 0.9191 & 0.9956 \\
		    AVIRIS Airport & 0.9526 & 0.9538 & 0.4807 & 0.9518 & 0.9928  \\
			AVIRIS Urban & 0.9907 &0.9910 & 0.9803 & 0.9860 & 0.9946 \\
			AVIRIS Beach & 0.9887 &0.9888  & 0.9917 & 0.9574 & 0.9926   \\
			\hline
		\end{tabular}
	\caption{AUC results for each method.}
	\label{tab:auc}
\end{table}

\begin{figure}[b]
\centering{
\setlength{\tabcolsep}{1pt}	
\begin{tabular}{ccc}
   \small  Real Image & \small Reconstructed Image & \small Difference Image  \\
        \includegraphics[width=2.8cm]{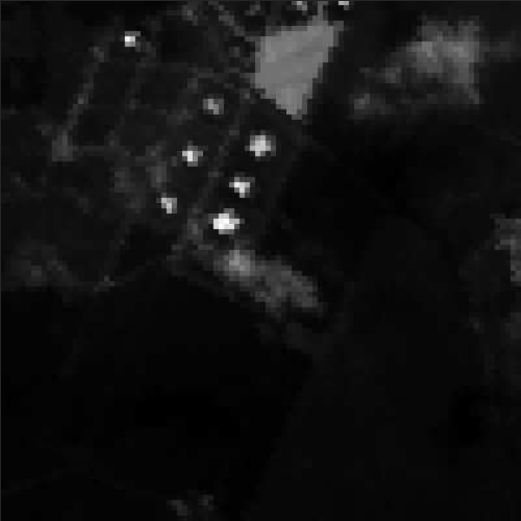} &
            \includegraphics[width=2.8cm]{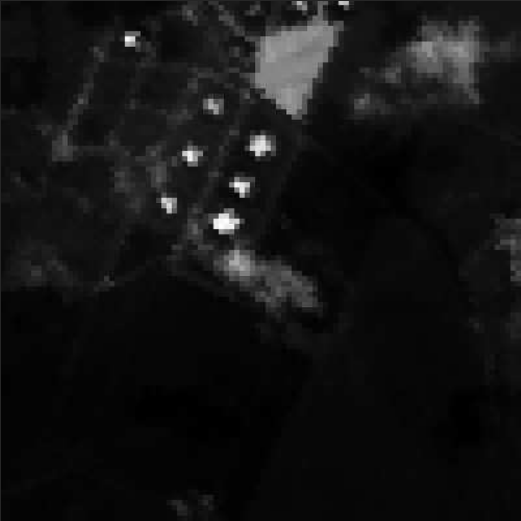} &
 			\includegraphics[width=2.8cm]{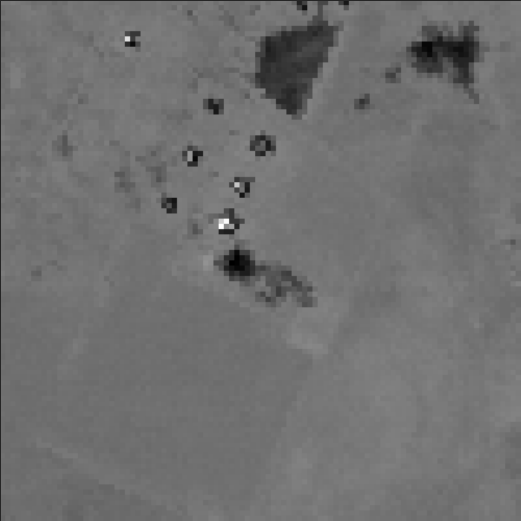} 

\end{tabular}

}
\caption{Real, reconstructed and spectral difference images for the 15th channel of AVIRIS Urban.}
\label{fig:diff}
\end{figure}

\begin{figure*}[t]
\centering{
\setlength{\tabcolsep}{1pt}	
\begin{tabular}{cccccccc}
   Input Image & Reference Map & RX & WRX & SVDD & DAEAD & GAN-RX  \\
   
    a)   \includegraphics[width=2.1cm]{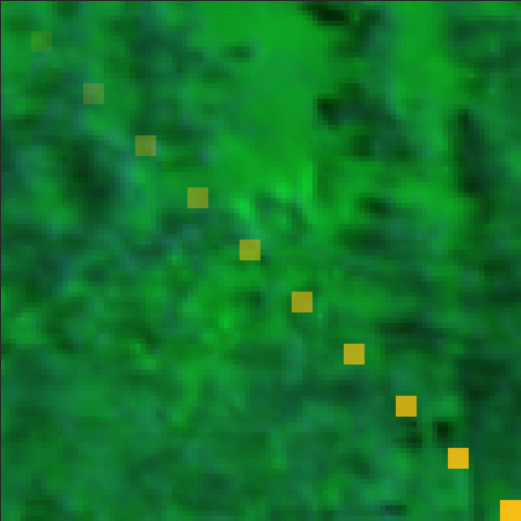} &
            \includegraphics[width=2.1cm]{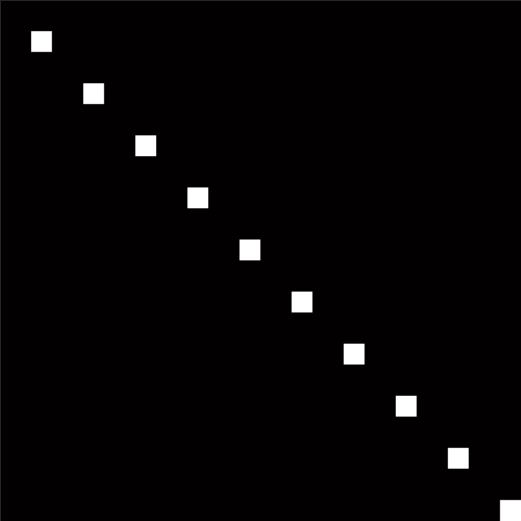} &
 			\includegraphics[width=2.1cm]{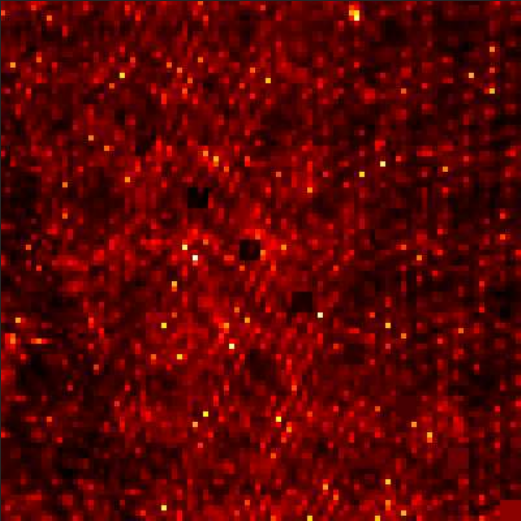}  &
 			\includegraphics[width=2.1cm]{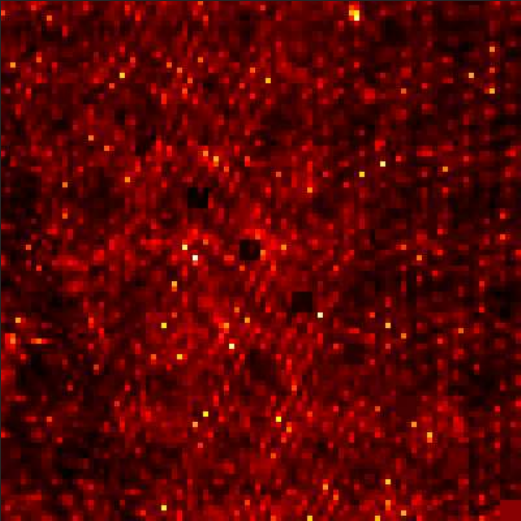}  &
 		    \includegraphics[width=2.1cm]{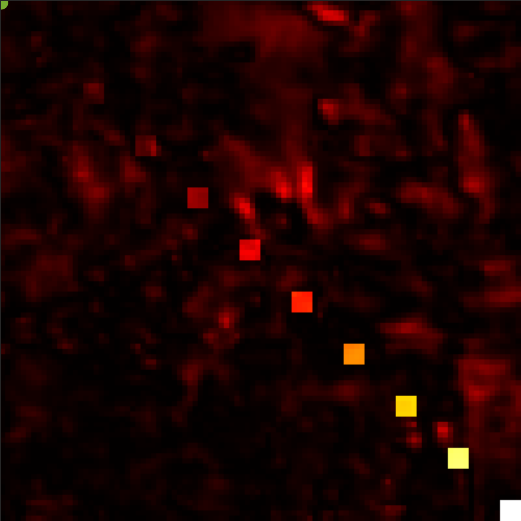} &
 		   \includegraphics[width=2.1cm]{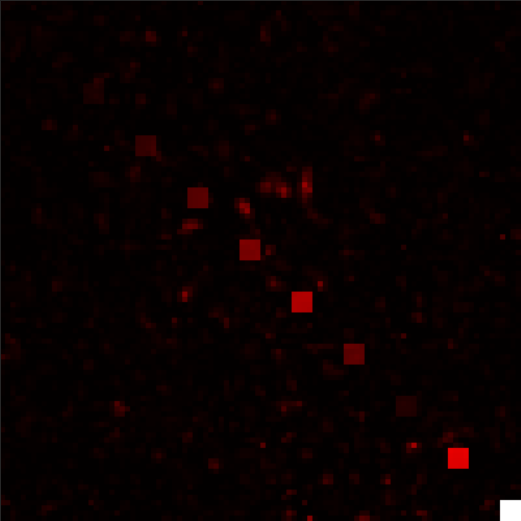} &
 			\includegraphics[width=2.1cm]{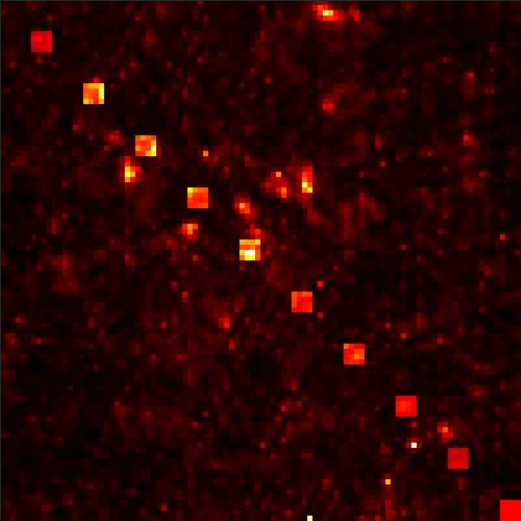}  \\
 			
 	b)   \includegraphics[width=2.1cm]{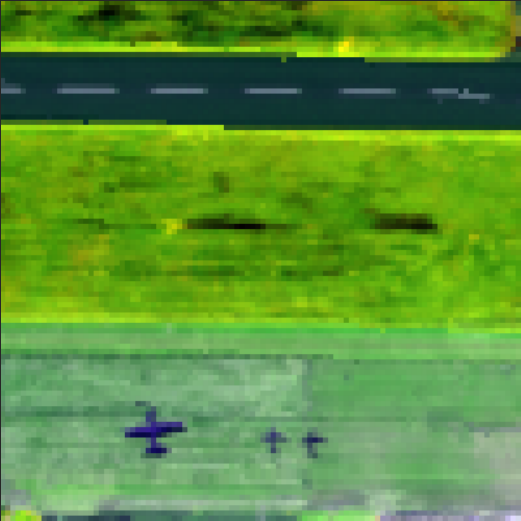} &
            \includegraphics[width=2.1cm]{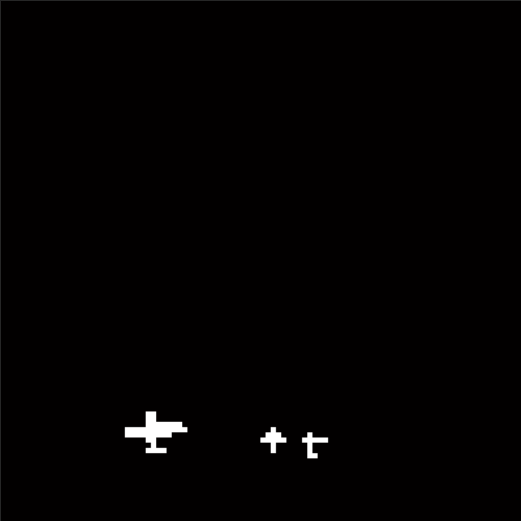} &
 			\includegraphics[width=2.1cm]{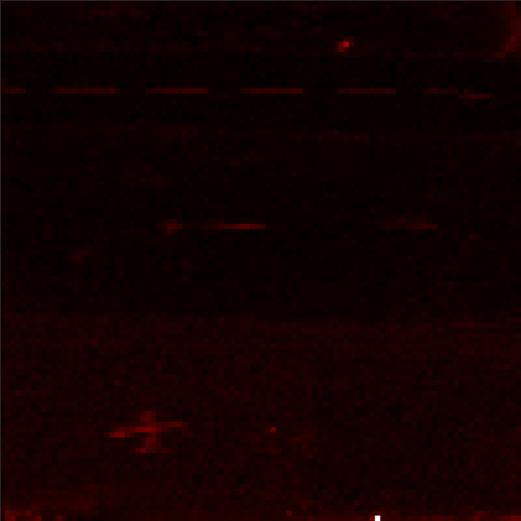}  &
 			\includegraphics[width=2.1cm]{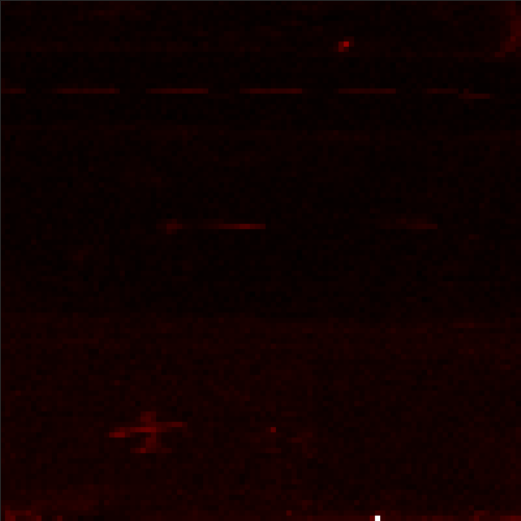}  &
 		    \includegraphics[width=2.1cm]{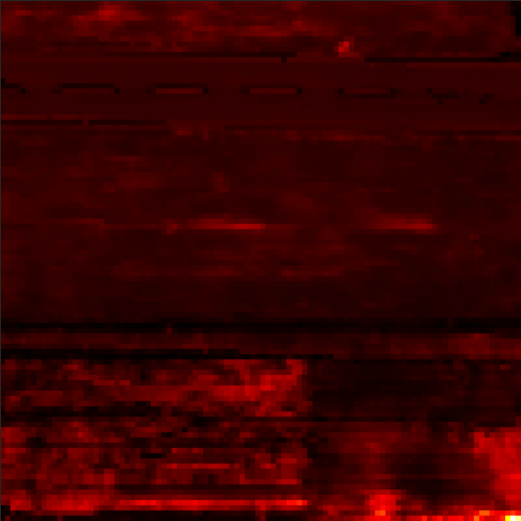} &
 		    \includegraphics[width=2.1cm]{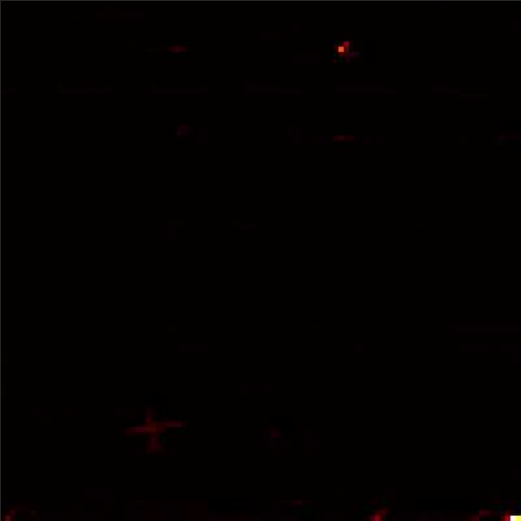} &
 			\includegraphics[width=2.1cm]{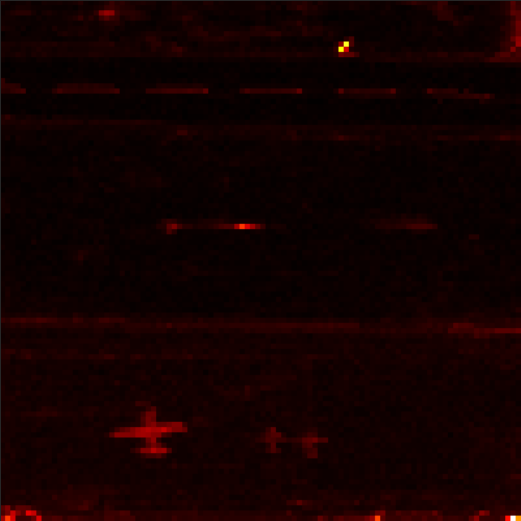}  \\

     c)   \includegraphics[width=2.1cm]{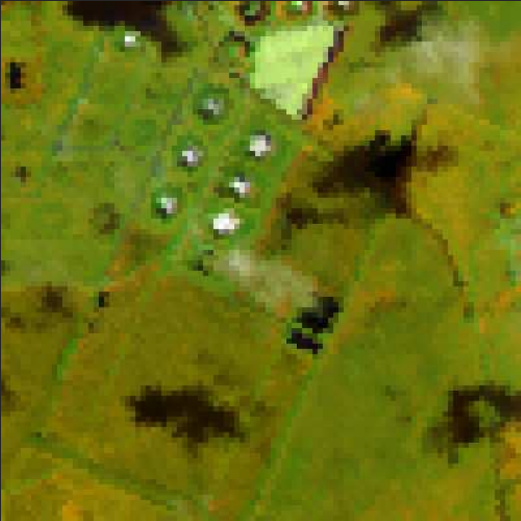} &
            \includegraphics[width=2.1cm]{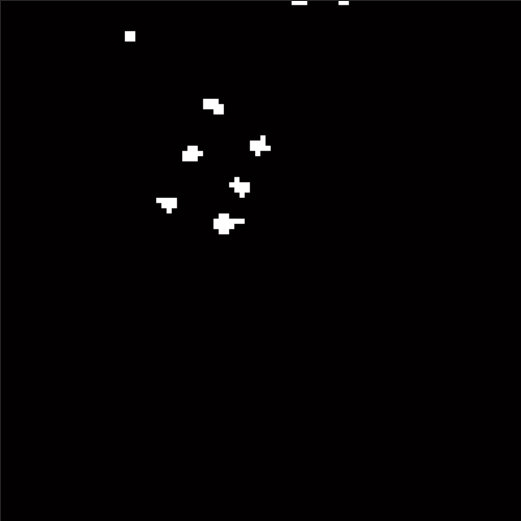} &
 			\includegraphics[width=2.1cm]{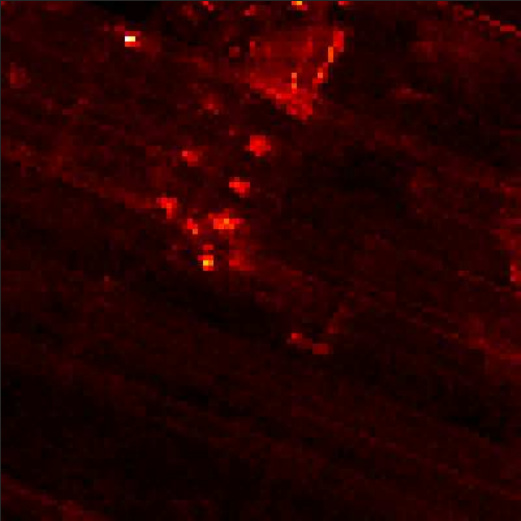}  &
 			\includegraphics[width=2.1cm]{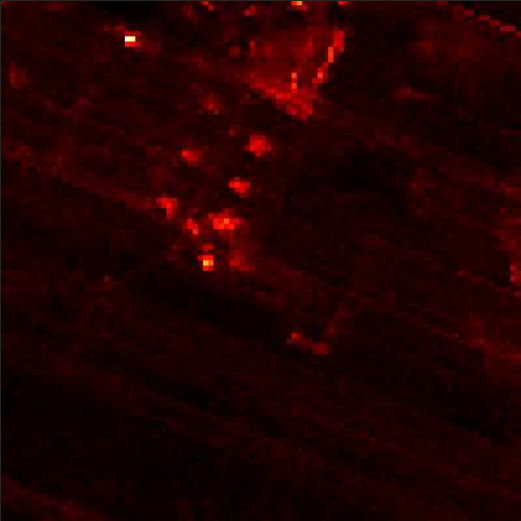}  &
 		    \includegraphics[width=2.1cm]{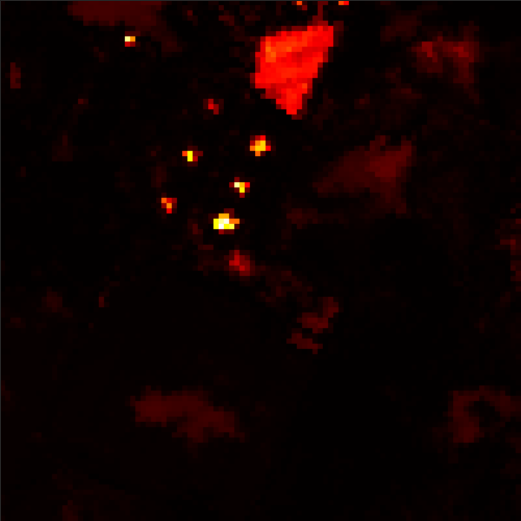} &
 		     \includegraphics[width=2.1cm]{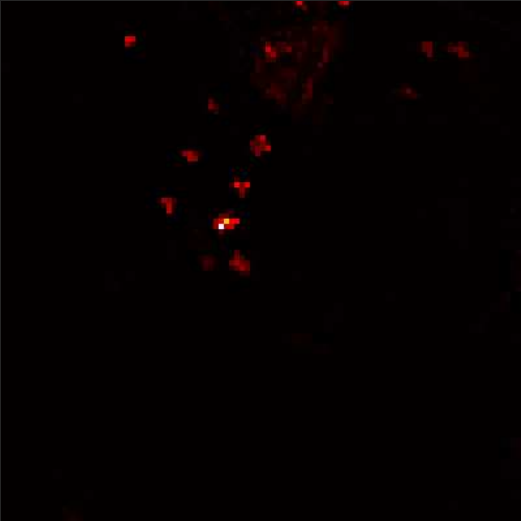}&
 			\includegraphics[width=2.1cm]{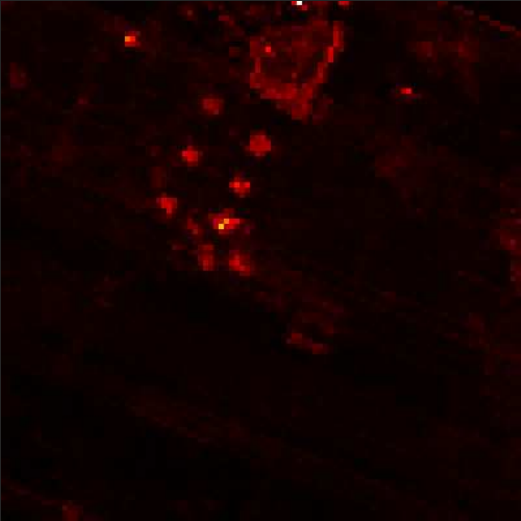}  \\
 			
 	 d)   \includegraphics[width=2.1cm]{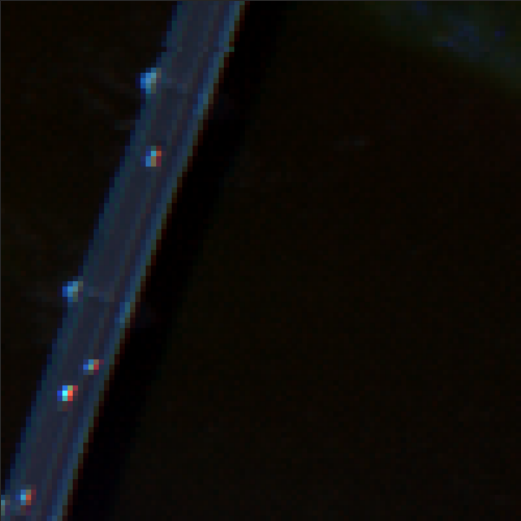} &
            \includegraphics[width=2.1cm]{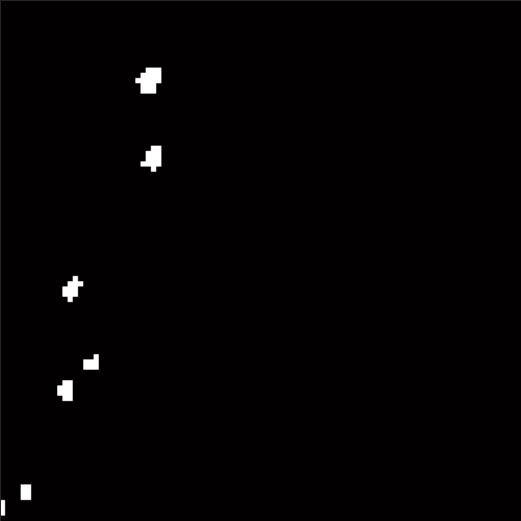} &
 			\includegraphics[width=2.1cm]{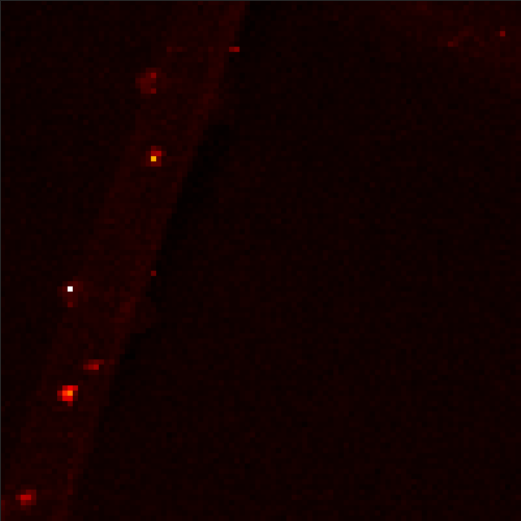}  &
 			\includegraphics[width=2.1cm]{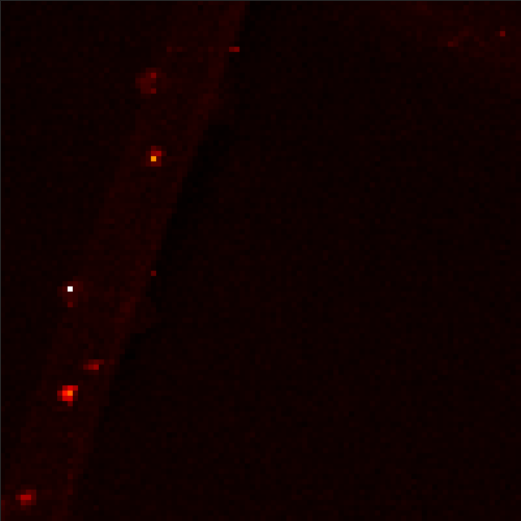}  &
 		    \includegraphics[width=2.1cm]{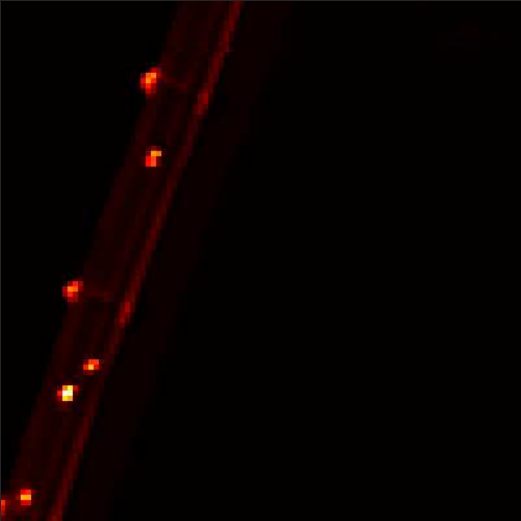} &
 		   \includegraphics[width=2.1cm]{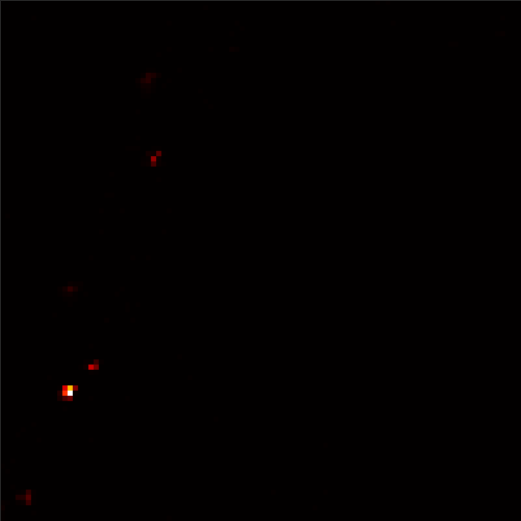} &
 			\includegraphics[width=2.1cm]{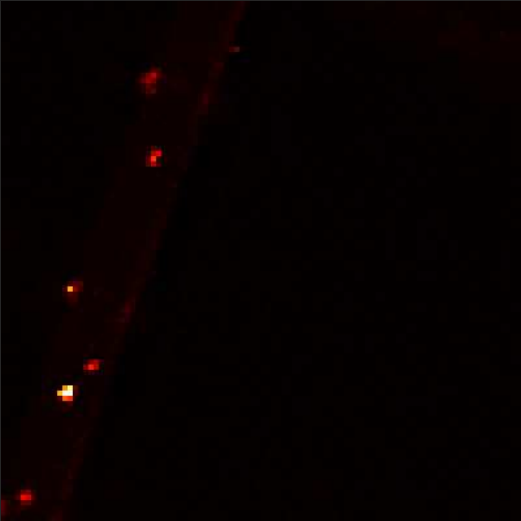}  \\

\end{tabular}

}
\caption{Three-band false color composite images, target reference maps and detection maps, a) Cooky City Synthetic, b) AVIRIS Airport, c) AVIRIS Urban and d) AVIRIS Beach.}
\label{fig:dm}
\end{figure*}

 \begin{figure}
  	\centering{
  	
\begin{tabular}{cc}
   a) Cooky City Synthetic & b) AVIRIS Airport \\
    \includegraphics[width=4.1cm]{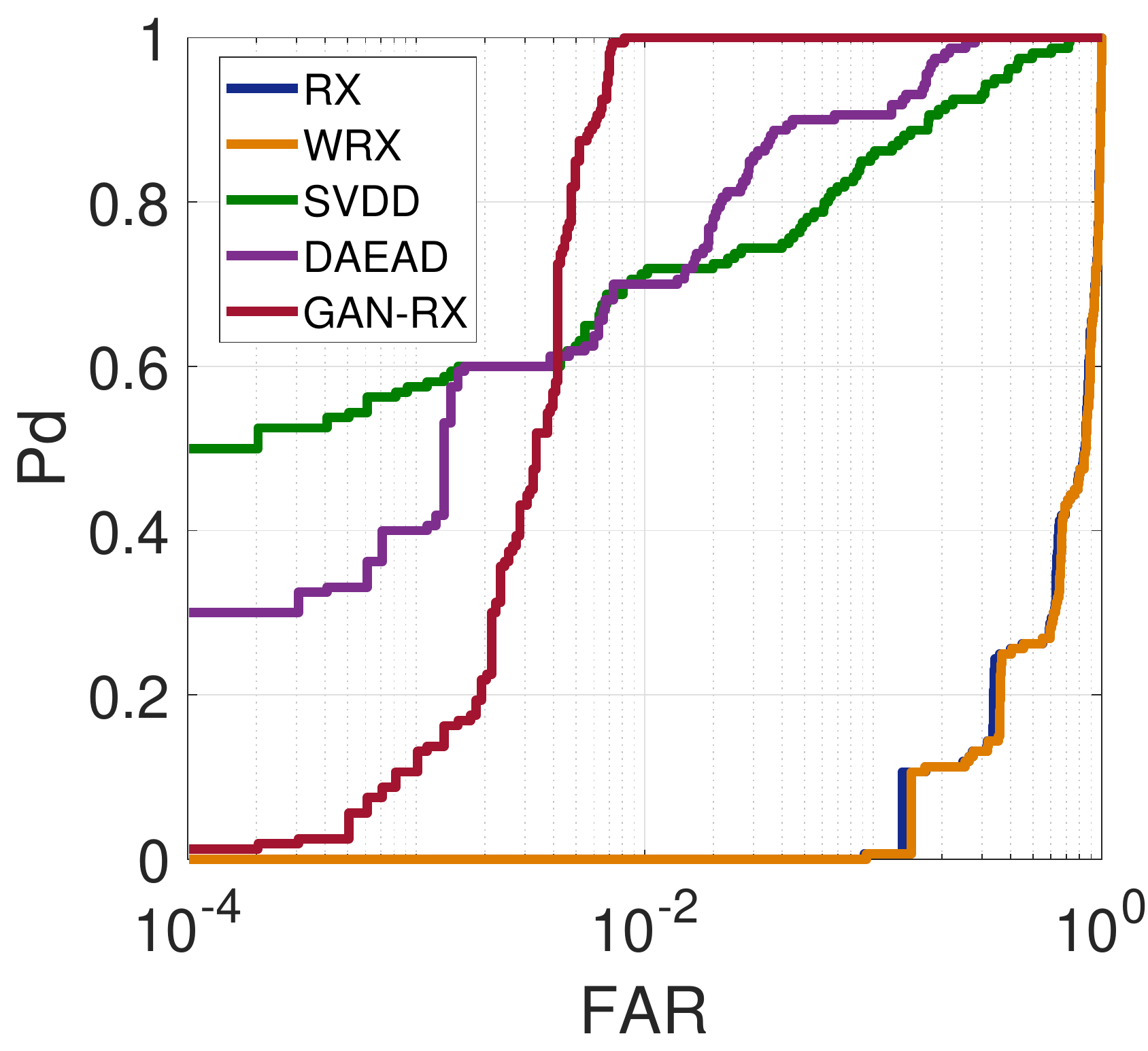} &
    \includegraphics[width=4.1cm]{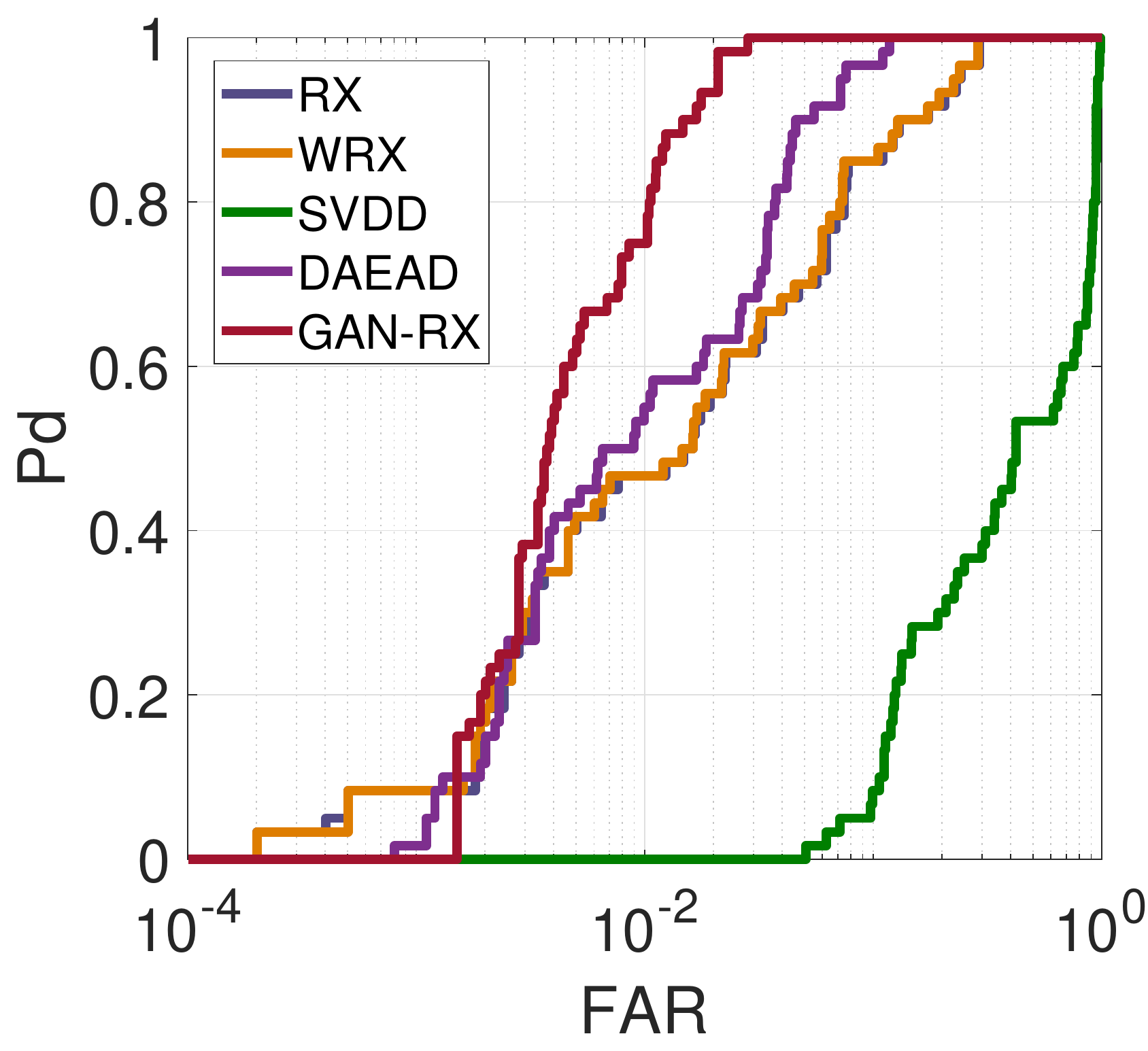}
 	  \\
 	   c) AVIRIS Urban & d) AVIRIS Beach \\
    \includegraphics[width=4.1cm]{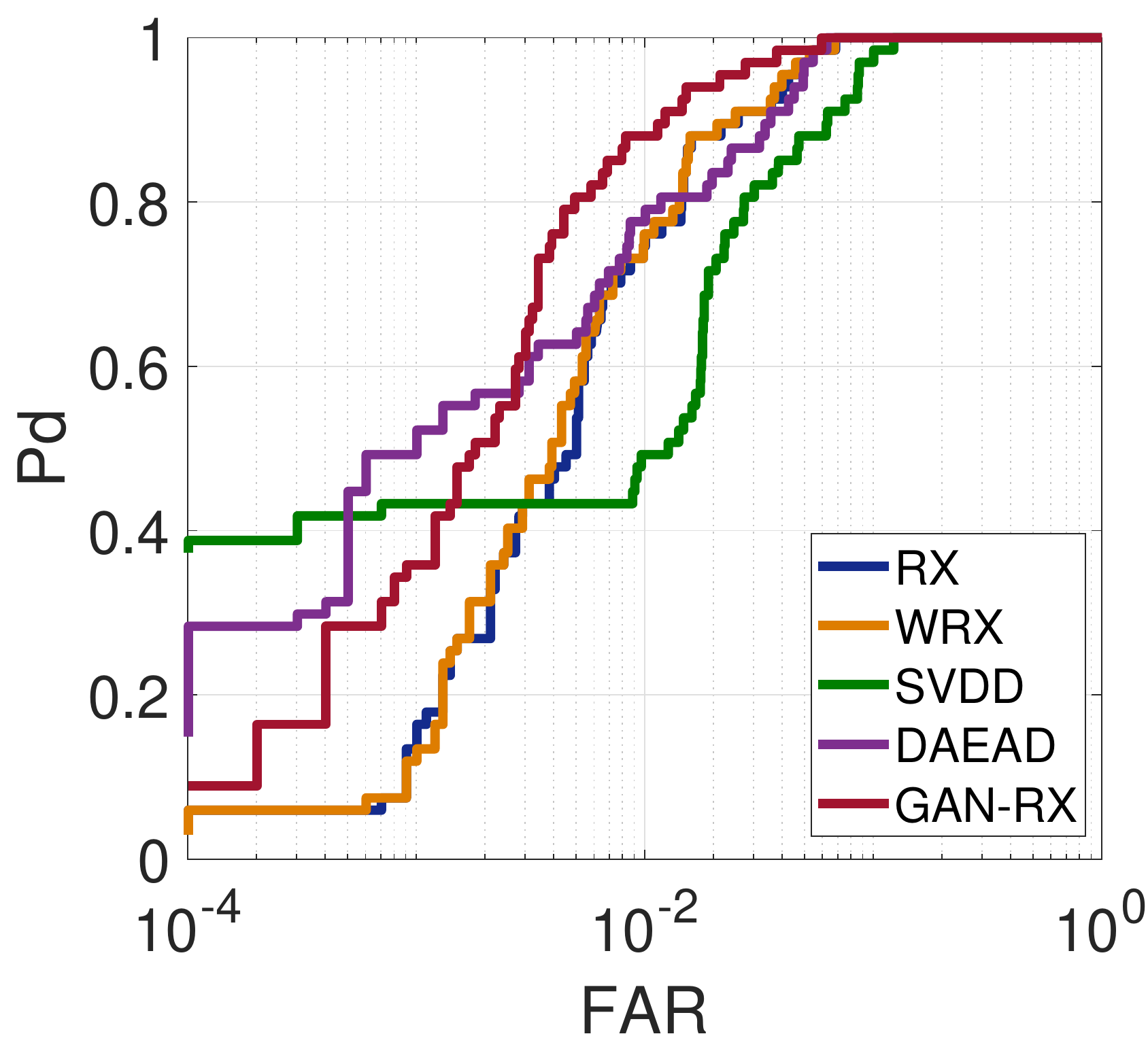} &
    \includegraphics[width=4.1cm]{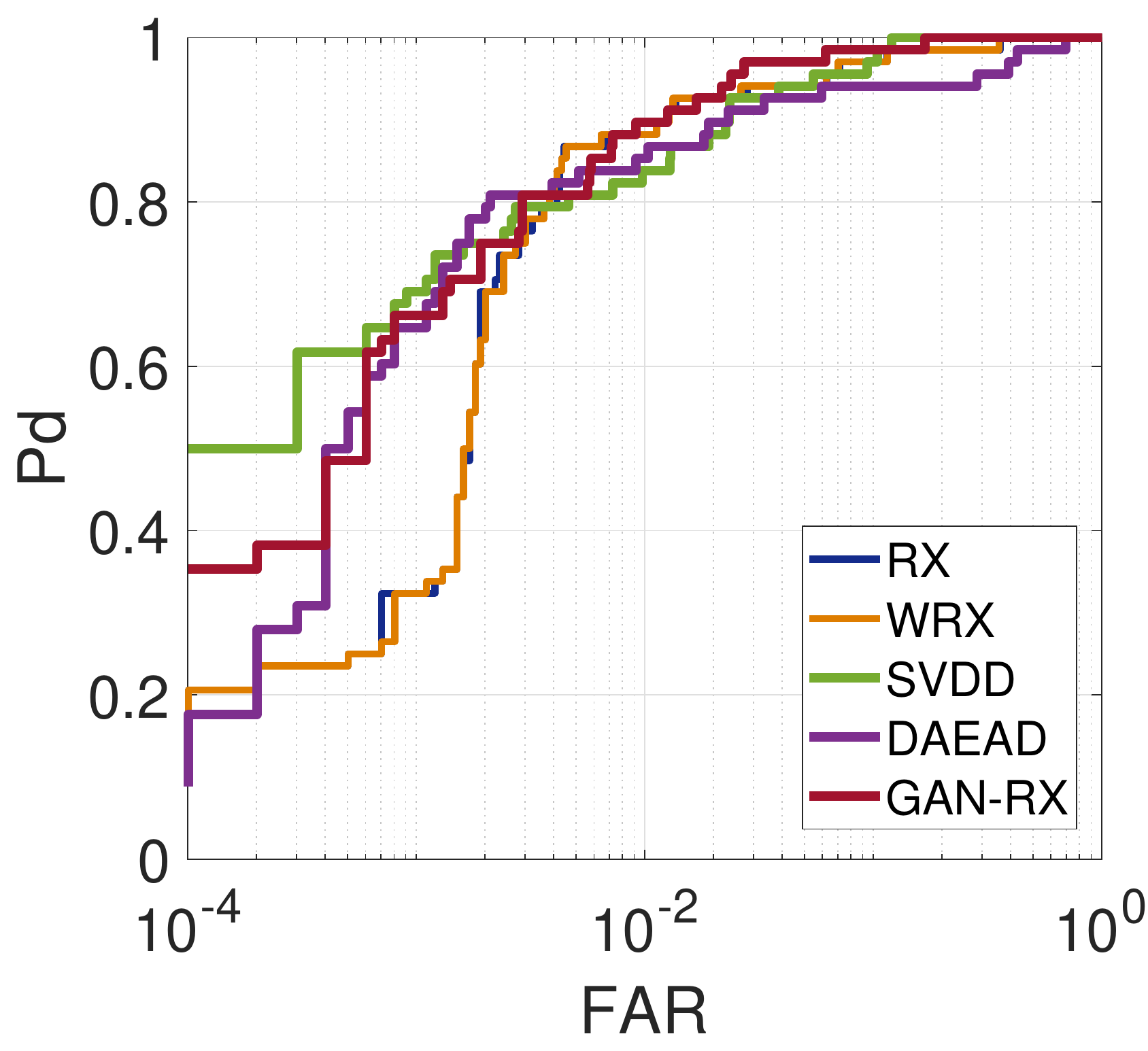}
\end{tabular}

}
\caption{ROC curves for each image.} \label{fig:roc}
\end{figure}

\subsection{Synthetic Data}

We use Cooky City hyperspectral image collected by the HyMap sensor over Cooky City, Montana, USA \cite{Snyder2008}. In this image, six different targets are located in different parts. We choose $100$ $\mathrm{x}$ $100$ vegetation region from the image and implant the car target called \textit{V2} in the region. The background and target spectral signatures are given in Fig. \ref{spectrumss}. As seen from this figure, the target signature is not close to the background signatures. We use linear mixture models to implant the target as follows:

\begin{eqnarray}\label{gengmm}
\mathbf{m}=a\mathbf{t}+(1-a)\mathbf{s},
\end{eqnarray}
where $a$ is abundance fraction of $\mathbf{t}$ target spectrum and $\mathbf{s}$ is background spectrum. We implant $4$ $\mathrm{x}$ $4$ targets in diagonal direction with different abundance fraction rates from 0.1 to 1. The image and the reference map are seen in the first and the second columns of Fig. \ref{fig:dm}.

\subsection{Real Data}

We use real hyperspectral images collected from AVIRIS sensor at different times. We choose different scene categories including airport, urban and beach areas. The size of the images is $100$ $\mathrm{x}$ $100$ and the number of spectral bands varies from $102$ to $204$. The detailed information can be found in \cite{kang2017hyperspectral}. The image and reference maps are given the first and the second columns of Fig. \ref{fig:dm}.

\subsection{Performance Evaluation}

We compare the detection performance of the algorithm using the ROC curve and AUC metric. We visually analyze the real, reconstructed and spectral difference images for the 15th channel of AVIRIS Urban image in Fig. \ref{fig:diff}. As seen from the difference image, the major part of the background is removed from the image and the anomaly pixels become more salient compared to the original image. In Fig. \ref{fig:roc}, the ROC curves are given for each test image. As seen from the figure, the proposed GAN-RX method provides higher detection probability with a low false alarm rate compared to other methods. In Table \ref{tab:auc}, we present AUC results for each method. Since the weights of the network depend on initial parameters, we run the GAN-RX and DAEAD methods 20 times and calculate the average AUC values. The performance of our proposed GAN-RX detector is the best one among the other methods.

In Fig. \ref{fig:dm}, we present the detection maps. As seen from the figure, in the GAN-RX detection maps, the anomaly pixels are more dominant and closer to the reference map than other detection maps.

\section{Conclusion}
\label{SecConc}
We proposed a GAN model for hyperspectral anomaly detection. The GAN model is trained to learn the background image. Then, we use the reconstructed image generated by the GAN model for background removal. In the experimental section, it is observed that the detection performance of the proposed GAN-RX is better than the classical RX and WRX anomaly detectors and deep autoencoder based DAEAD method. As a future study, rather than using an RX-AD, we plan to test different detectors on the spectral difference image to improve the detection performance. 

\bibliographystyle{IEEEtran}

\end{document}